\documentclass[aps,prl,twocolumn,groupedaddress]{revtex4}

\begin{document}

\newcommand{\Ui}{U^\prime}
\newcommand{\Uii}{U^{\prime\prime}}
\newcommand{\Uiii}{U^{\prime\prime\prime}}
\newcommand{\Uiiii}{U^{\prime\prime\prime\prime}}

\newcommand{\Z}{Z_0}
\newcommand{\Zi}{{Z_0^\prime}}
\newcommand{\Zii}{Z_0^{\prime\prime}}

\newcommand{\ZZ}{Z_1}
\newcommand{\ZZi}{{Z_1^\prime}}
\newcommand{\ZZii}{Z_1^{\prime\prime}}

\newcommand{\ui}{u^\prime}
\newcommand{\uii}{u^{\prime\prime}}
\newcommand{\uiii}{u^{\prime\prime\prime}}
\newcommand{\uiiii}{u^{\prime\prime\prime\prime}}

\newcommand{\z}{z_0}
\newcommand{\zi}{{z_0^\prime}}
\newcommand{\zii}{z_0^{\prime\prime}}

\newcommand{\zz}{z_1}
\newcommand{\zzi}{{z_1^\prime}}
\newcommand{\zzii}{z_1^{\prime\prime}}

\title{The Effective Average Action Beyond First Order}
\author{H. Ballhausen}
\email{ballhausen@physi.uni-heidelberg.de}
\affiliation{Institute for Theoretical Physics, Heidelberg University\\Philosophenweg 16, 69120 Heidelberg, Germany}

\begin{abstract}
\noindent
A derivative expansion of the effective average action beyond first order yields renormalization
group functional flow equations which are used for the computation of critical exponents of the
Ising universality class. The critical exponent $\nu$ in $D=3$ is consistent with high-precision methods.
\end{abstract}

\maketitle

\subsection*{The effective average action}

\noindent
The effective average action $\Gamma_k[\varphi]$ ( see \cite{Lit1} and \cite{Lit2,Lit3,Lit4,Lit5} ) is a functional of the fields
$\varphi$. It interpolates between the classical action $S$ at some microscopic ultraviolet scale
$\Lambda$ and the effective action $\Gamma$: $\Gamma_\Lambda=S$, $\lim_{k\to 0} \Gamma_k = \Gamma$.
Its dependence on the momentum scale $k$ is described by an exact renormalization group
functional flow equation:

\bigskip $ \partial_k \Gamma_k = \frac{1}{2} \textrm{Tr} \{ (\Gamma_k^{(2)} + R_k)^{-1} \partial_k R_k \} $

\bigskip
\noindent
where $\Gamma_k^{(2)}$ is the second functional derivative of $\Gamma_k$ with respect to the
fields and $R_k$ denotes a momentum cutoff function with the properties $\lim_{k\to 0}R_k=0$,
$\lim_{k\to \Lambda}R_k \to \infty$ and $\lim_{q^2\to 0}R_k(q^2) > 0$ ensuring the above limit
properties of $\Gamma_k$. Universal properties of the renormalization group flow, however, are
independent of the actual choice of $R_k$.

\bigskip
\noindent
The flow of $\Gamma_k^{(2)}$ and higher n-point-functions can be derived from the flow of
$\Gamma_k$ by functional derivation, e.g.

\bigskip $ \partial_k \Gamma_k^{(2)} = \textrm{Tr} \{ \Gamma_k^{(3)2} (\Gamma_k^{(2)} + R_k)^{-3} \partial_k R_k \} $ \\
$ ~~~~~~~~~~~~ - \frac{1}{2} \textrm{Tr} \{ \Gamma_k^{(4)} (\Gamma_k^{(2)} + R_k)^{-2} \partial_k R_k \} $

\bigskip
\noindent
where the flow of $\Gamma_k^{(3)}$ would involve $\Gamma_k^{(4)}$ and $\Gamma_k^{(5)}$ and so on.
In order to end up with a closed system of functional flow equations one needs to do a finite
truncation of $\Gamma_k$.

\bigskip
\noindent
In the following we concentrate on a one-component real scalar field $\varphi: I\!\!R^D \to
I\!\!R$. There are the invariants $\rho=\frac{1}{2}\varphi\varphi$ and
$\rho q^2=\frac{1}{2}\partial_\mu\varphi\partial^\mu\varphi$ featuring the $O(1)$ symmetry of
the Ising universality class. Here the lowest order truncation of the effective average action
reads

\bigskip $ \Gamma_k = \int d^Dx \{ U_k(\rho) + \frac{1}{2}\partial_\mu\varphi\partial^\mu\varphi Z_k + O(\partial^2) \} $

\bigskip
\noindent
in terms of the effective average potential $ U_k(\rho) $ and a field independent
wave function renormalization $ Z_k $. The effective average propagator

\bigskip $ \Gamma_k^{(2)}(q^2) + R_k(q^2) = U_k^\prime(\rho) + 2 \rho U_k^{\prime\prime}(\rho) + q^2 Z_k + R_k(q^2) $

\bigskip
\noindent
gives the flow of the effective average potential

\bigskip $ \partial_k U_k(\rho) = \frac{1}{2} \int \frac{d^Dq}{(2\pi)^D} \frac{\partial_k R_k}{U_k^\prime(\rho) + 2 \rho U_k^{\prime\prime}(\rho) + q^2 Z_k + R_k(q^2)} $

\bigskip
\noindent
and $ \partial_k Z_k $ can be determined from $ \partial_{q^2} \partial_k \Gamma_k^{(2)}(q^2) \mid_{q^2=0} $.

\newpage
\subsection*{Derivative expansion beyond first order}

\noindent
In a second order derivative expansion of the effective average action there are
contributions from up to four nonvanishing momenta. These are parametrized by the
effective average potential and one respectively three linearly independent wave
function renormalizations in first respectively second order:

\bigskip $ \Gamma_k = \int \limits d^Dx \{ U_k(\rho) $\\
$ ~~~~~~~~~~~~~~\; + Z_{A,k}(\rho) \cdot \frac{1}{2!} \partial_\mu \varphi \partial^\mu \varphi $\\
$ ~~~~~~~~~~~~~~\; + Z_{B,k}(\rho) \cdot \frac{1}{4!} \partial_\mu \varphi \partial^\mu \varphi \partial_\nu \varphi \partial^\nu \varphi $\\
$ ~~~~~~~~~~~~~~\; + Z_{C,k}(\rho) \cdot \frac{1}{4!} \varphi \partial_\mu \partial^\mu \varphi \partial_\nu \varphi \partial^\nu \varphi $\\
$ ~~~~~~~~~~~~~~\; + Z_{D,k}(\rho) \cdot \frac{1}{4!} \varphi \varphi \partial_\mu \partial^\mu \varphi \partial_\nu \partial^\nu \varphi + O(\partial^6) \} $

\bigskip
\noindent
This truncation has been the starting point for a recent investigation in \cite{Lit5a}. There an additional field expansion of the potential and 
the wave function renormalizations in powers of $\rho$ is done. Here we keep the full field dependence but instead go 
only one step beyond first order and take only $U_k$, $Z_{A,k}$ and $Z_{D,k}$ into account due to the following reasoning:

\bigskip
\noindent
While $Z_{A,k}$ and $Z_{D,k}$ directly contribute to $ \Gamma_k^{(2)} $ and thus enter all flow 
equations as $q^2$ and $q^4$ terms in the denominators of the integrands, $Z_{B,k}$ and $Z_{C,k}$
only give additive contributions to the four-point-vertex and the three-point-vertex respectively. 
Also, their flow requires $\partial_k \Gamma_k^{(4)}$ and $\partial_k \Gamma_k^{(3)}$. 
This is why we neglect $Z_{B,k}$ and $Z_{C,k}$ in this paper which corresponds to a truncation 
beyond first and below second order.

\bigskip
\noindent
We hope that this improves the critical exponent $\nu$ compared to first order calculations.
On the other hand, the anomalous dimension $\eta$ is related to the momentum dependence of the
wave function renormalization and will probably suffer from the missing contributions.

\bigskip
\noindent
Substituting $U:=U_k$, $\Z:=Z_{A,k}$ and $\ZZ:=\frac{2!}{4!}\varphi \varphi Z_{D,k}$ the 
truncation thus reads:

\bigskip $ \Gamma_k = \int \limits d^Dx \{ U(\rho) $\\
$ ~~~~~~~~~~~~~~~~ + \Z(\rho) \cdot \frac{1}{2!} \partial_\mu \varphi \partial^\mu \varphi $\\
$ ~~~~~~~~~~~~~~~~ + \ZZ(\rho) \cdot \frac{1}{2!} \partial_\mu \partial^\mu \varphi \partial_\nu \partial^\nu \varphi + O(\partial^4) \} $

\bigskip
\noindent
and the next step is to derive $\partial_k U$, $\partial_k \Z$ and $\partial_k \ZZ$:

\newpage
\subsection*{Renormalization group flow equations}

\noindent
Expanding the fields $\varphi(x)=\bar{\varphi}+\chi(x)$ in small fluctuations
$\chi(x) = \sum_q e^{iqx} \chi(q)$ in momentum space around a constant background 
$\bar{\varphi}$ yields the n-point-functions
$ \Gamma^{(n)}(q_1,...,q_n) = \frac{\delta}{\delta \varphi(q_1)} \cdot\!\cdot\!\cdot \frac{\delta}{\delta \varphi(q_n)} \; \Gamma = \frac{\delta}{\delta \chi(q_1)} \cdot\!\cdot\!\cdot \frac{\delta}{\delta \chi(q_n)} \; \Gamma \mid_{\chi=0} $:

\bigskip $ \Gamma^{(2)}(q,-q) = \Ui + 2 \rho \Uii + q^2 \cdot \Z + q^4 \cdot \ZZ $

\smallskip $ \Gamma^{(2)}(p,-p) = \Ui + 2 \rho \Uii + p^2 \cdot \Z + p^4 \cdot \ZZ $

\smallskip $ \Gamma^{(3)}(p,q,-p-q) = $\\
$ ~~~~ \Gamma^{(3)}(p+q,-q,-p) = \sqrt{2\rho} ~ \{ ~ 3 \Uii + 2 \rho \Uiii $\\
$ ~~~~~~~~~ + (p^2+(pq)+q^2) \cdot \Zi $\\
$ ~~~~~~~~~ + (p^4+2p^2(pq)+3p^2q^2+2(pq)q^2+q^4) \cdot \ZZi ~ \} $

\smallskip $ \Gamma^{(4)}(p,q,-q,-p) = 3 \Uii + 12 \rho \Uiii + 4 \rho^2 \Uiiii $\\
$ ~~~~~~~~~ +(p^2+q^2) \cdot ( \Zi + 2 \rho \Zii ) $\\
$ ~~~~~~~~~ +(p^4+4p^2q^2+q^4) \cdot ( \ZZi + 2 \rho \ZZii ) $

\bigskip
\noindent
These are inserted into

\bigskip $ \partial_k \Gamma = \frac{1}{2} \int \frac{d^Dp}{(2\pi)^D} \frac{\partial_k R(p^2)}{\Gamma^{(2)}(p,-p) + R(p^2)} $

\smallskip $ \partial_k \Gamma^{(2)}(q,-q) $\\
$ ~~~~~~~\; = \int \frac{d^Dp}{(2\pi)^D} \frac{\Gamma^{(3)}(p,q,-p-q)\Gamma^{(3)}(p+q,-q,-p) ~ \partial_k R(p^2)}{(\Gamma^{(2)}(p^2)+R(p^2))^2(\Gamma^{(2)}((p+q)^2)+R((p+q)^2))} $\\
$ ~~~~~~~\; - \frac{1}{2} \int \frac{d^Dp}{(2\pi)^D} \frac{\Gamma^{(4)}(p,q,-q,-p) ~ \partial_k R(p^2)}{(\Gamma^{(2)}(p^2)+R(p^2))^2} $

\bigskip
\noindent
and yield the flow equations

\bigskip $ \partial_k U = \partial_k \Gamma $

\smallskip $ \partial_k \Z = \frac{\partial}{\partial q^2} ~ \partial_k \Gamma^{(2)} ~ \mid_{q^2=0} $

\smallskip $ \partial_k \ZZ = \frac{1}{2} \frac{\partial}{\partial q^2} \frac{\partial}{\partial q^2} ~ \partial_k \Gamma^{(2)} ~ \mid_{q^2=0} $

\bigskip
\noindent
In order to be able to perform the $q^2$-derivatives one has to expand the denominator
containing $(p+q)^2$. Let $N=\Gamma^{(2)}(p^2)+R(p^2)$ and $M=\Gamma^{(2)}((p+q)^2)+R((p+q)^2)$:

\bigskip $ \frac{1}{M} = \frac{1}{N} - \frac{(2pq+q^2)\dot{N}}{N^2} + \frac{(2pq+q^2)^2\dot{N}^2}{N^3} - \frac{1}{2}\frac{(2pq+q^2)^2\ddot{N}}{N^2} + ... $

\bigskip
\noindent
Then the mixed scalar products $(pq)$ can also be eliminated:

\bigskip $ \frac{1}{2} \int \frac{d^Dp}{(2\pi)^D} f(p^2,q^2) \cdot (pq)^2 = \frac{1}{2} \int \frac{d^Dp}{(2\pi)^D} f(p^2,q^2) \frac{p^2q^2}{D} $

\smallskip $ \frac{1}{2} \int \frac{d^Dp}{(2\pi)^D} f(p^2,q^2) \cdot (pq)^4 = \frac{1}{2} \int \frac{d^Dp}{(2\pi)^D} f(p^2,q^2) \frac{p^4q^4}{D(D+2)} $

\smallskip $ \frac{1}{2} \int \frac{d^Dp}{(2\pi)^D} f(p^2,q^2) \cdot (pq)^k = 0 $ for all odd $k$

\bigskip
\noindent
This finally gives the explicit flow equations:

\bigskip $ \partial_k U = 1 \cdot I\Big[ \frac{1}{N} \Big] $

\smallskip $ \partial_k \Z = 4 \rho C_0^2 \cdot I\Big[ - \frac{\dot{N}}{N^4} + \frac{4}{D} \frac{p^2\dot{N}^2}{N^5} - \frac{2}{D} \frac{p^2\ddot{N}}{N^4} \Big] + ... $

\smallskip $ \partial_k \ZZ = 2 \rho C_0^2 \cdot I\Big[ \frac{2\dot{N}^2}{N^5} - \frac{\ddot{N}}{N^4} \Big] + ... $

\bigskip $ I\Big[f(p^2)\Big] = \frac{1}{2} \int \frac{d^Dp}{(2\pi)^D} ~ f(p^2) ~ \partial_k R(p^2) $

\smallskip $ C_0 = 3\Uii+2\rho\Uiii $

\newpage
\subsection*{Scale invariant flow equations}

\noindent
As we are interested in critical phenomena of second order phase transitions 
corresponding to fix points in the renormalization group flow, we write the
flow equations in explicitely scale invariant form.

\bigskip
\noindent
To this end one substitutes $ \z(\sigma) = \Z(\rho) / \Z(\rho_k) $ ( where
$\rho_k$ is the running minimum of $U(\rho)$ ), $\sigma=\Z(\rho_k)k^{2-D}\rho$,
$ u = k^{-d} U $, $t=-\textrm{Ln}k$, $\eta=-\partial_t \textrm{Ln}\Z(\rho_k)$ etc.
such that

\bigskip $ \partial_t u^\prime = (-2+\eta) u^\prime +(D-2+\eta)\sigma\uii $\\
$ ~~~~ - c_0 \cdot I_2^0 $\\
$ ~~~~ - c_1 \cdot I_2^1 $

\bigskip $ \partial_t \z = \eta z + (D-2+\eta) \sigma \zi $\\
$ ~~~~ + 4 \sigma c_0 c_0 \cdot ( ~~~~~~~~~~           - \frac{~0+~D}{D} J_4^0 + \frac{2}{D} K_4^1 ) $\\
$ ~~~~ + 8 \sigma c_0 c_1 \cdot ( \frac{0+1D}{D} I_3^0 - \frac{~2+~D}{D} J_4^1 + \frac{2}{D} K_4^2 ) $\\
$ ~~~~ + 4 \sigma c_1 c_1 \cdot ( \frac{1+2D}{D} I_3^1 - \frac{~4+~D}{D} J_4^2 + \frac{2}{D} K_4^3 ) $\\
$ ~~~~ + 8 \sigma c_0 c_2 \cdot ( \frac{0+3D}{D} I_3^1 - \frac{~4+~D}{D} J_4^2 + \frac{2}{D} K_4^3 ) $\\
$ ~~~~ + 4 \sigma c_1 c_2 \cdot ( \frac{4+8D}{D} I_3^2 - \frac{12+2D}{D} J_4^3 + \frac{4}{D} K_4^4 ) $\\
$ ~~~~ + 4 \sigma c_2 c_2 \cdot ( \frac{4+6D}{D} I_3^3 - \frac{~8+~D}{D} J_4^4 + \frac{2}{D} K_4^5 ) $\\
$ ~~~~ - c_4 \cdot I_2^0 $\\
$ ~~~~ - c_5 \cdot 4 I_2^1 $

\bigskip $ \partial_t \zz = (2+\eta) \zz + (D-2+\eta) \sigma \zz^\prime $\\
$ ~~~~ + 2 \sigma c_0 c_0 \cdot ( ~~~~~~~~~~~~~~~~~~~~~~~~~~~~~~~~~~~~~~~~~~~~~\;             K_4^0 ) $\\
$ ~~~~ + 4 \sigma c_0 c_1 \cdot ( ~~~~~~~~~~~~            - \frac{~0+~2D}{D} J_4^0 + \frac{~~~~~8+D~~~\;}{D} K_4^1 ) $\\
$ ~~~~ + 2 \sigma c_1 c_1 \cdot ( \frac{~0+~2D}{D} I_3^0 - \frac{10+~4D}{D} J_4^1 + \frac{~~~~20+D~~~\,}{D} K_4^2 ) $\\
$ ~~~~ + 4 \sigma c_0 c_2 \cdot ( \frac{~0+~2D}{D} I_3^0 - \frac{~8+~6D}{D} J_4^1 + \frac{36+18D+D^2}{D(D+2)} K_4^2 ) $\\
$ ~~~~ + 4 \sigma c_1 c_2 \cdot ( \frac{~4+~8D}{D} I_3^1 - \frac{32+~8D}{D} J_4^2 + \frac{64+30D+D^2}{D(D+2)} K_4^3 ) $\\
$ ~~~~ + 2 \sigma c_2 c_2 \cdot ( \frac{16+22D}{D} I_3^2 - \frac{72+12D}{D} J_4^3 + \frac{96+42D+D^2}{D(D+2)} K_4^4 ) $\\
$ ~~~~ - c_5 \cdot I_2^0 $

\bigskip
\noindent
where for abbreviation we have used the constants

\bigskip $ w = \ui + 2 \sigma \uii $\\
$ ~~~~ c_0 = w^\prime $\\
$ ~~~~ c_1 = \zi $\\
$ ~~~~ c_2 = \zzi $\\
$ ~~~~ c_3 = w^\prime + 2 \sigma w^{\prime\prime} $\\
$ ~~~~ c_4 = \zi + 2 \sigma \zii $\\
$ ~~~~ c_5 = \zzi + 2 \sigma \zzii $

\bigskip
\noindent
and the momentum integrals

\bigskip $ I^m_n(w,\z,\zz,\eta)[r] := \frac{1}{2} \int \frac{d^Dp}{(2\pi)^D} \frac{(p^2)^m ~ \tilde{\partial}_t r}{(w+p^2\z+p^4\zz+r)^n} $

\smallskip $ J^m_n(w,\z,\zz,\eta)[r] := \frac{1}{2} \int \frac{d^Dp}{(2\pi)^D} \frac{(p^2)^m ~ (\z+2 p^2 \zz+\dot{r}) ~ \tilde{\partial}_t r}{(w+p^2\z+p^4\zz+r)^n} $

\smallskip $ K^m_n(w,\z,\zz,\eta)[r] := \frac{1}{2} \int \frac{d^Dp}{(2\pi)^D} \frac{(p^2)^m ~ 2(\z+2 p^2 \zz+\dot{r})^2 ~ \tilde{\partial}_t r}{(w+p^2\z+p^4\zz+r)^{n+1}} $\\
$ ~~~~~~~~~~~~~~~~~~~~~~~~~~~~~ - \frac{1}{2} \int \frac{d^Dp}{(2\pi)^D} \frac{(p^2)^m ~ (2\zz + \ddot{r}) ~ \tilde{\partial}_t r}{(w+p^2\z+p^4\zz+r)^n} $

\bigskip
\noindent
which will be discussed next:

\newpage
\subsection*{Momentum integrals}

\noindent
The momentum integrals $I$, $J$ and $K$ can be computed efficiently in case of 
a linear cutoff. For the choice \cite{Lit6,Lit6a,Lit6b,Lit6c,Lit6d} 
$R_k=Z_k(k^2-q^2)\Theta(k^2-q^2)$ one has

\bigskip $ r = (1-p^2)\cdot\Theta(1-p^2) $

\medskip $ \dot{r} = -\Theta(1-p^2)-(1-p^2)\cdot\delta(1-p^2) $

\medskip $ \ddot{r} = 2\,\delta(1-p^2)+(1-p^2)\cdot\delta^\prime(1-p^2)$

\medskip $ \tilde{\partial}_t r = \frac{k\partial_k R_k}{Z_k k^2} = (2-\eta+p^2\eta)\cdot\Theta(1-p^2) $

\bigskip
\noindent
and $I$, $J$ and $K$ are reduced to a single integral $H$:

\bigskip $ I^m_n = (2-\eta) H_n^m $

\smallskip $ ~~~~~\,  + \eta H_n^{m+1} $

\medskip $ J^m_n = (\z-1) I^m_n $

\smallskip $ ~~~~~\,  + 2\zz I^{m+1}_n $

\medskip $ K^m_n = 2(\z-1)^2 I^m_{n+1} $

\smallskip $ ~~~~~\,  + 8(\z-1)\zz I^{m+1}_{n+1} $

\smallskip $ ~~~~~\,  + 8\zz^2 I^{m+2}_{n+1} $

\smallskip $ ~~~~~\, - 2\zz I^m_n $

\smallskip $ ~~~~~\, - v_D (w+\z+\zz)^{-n} $

\bigskip
\noindent
where $H$ is a special case of a hypergeometric function

\bigskip $ H^m_n(w,\z,\zz) $

\medskip $ = \frac{1}{2} \int \frac{d^Dp}{(2\pi)^D} \frac{(p^2)^m ~ \Theta(1-p^2)}{(w+p^2\z+p^4\zz+r)^n} $

\smallskip $ = v_D \int \limits_0^1 dx \frac{x^a}{((w+1)+(\z-1)x+\zz x^2)^n} $

\smallskip $ = \frac{v_D}{(w+1)^n} \int \limits_0^1 dx \cdot x^a \cdot (1-bx-cx^2)^{-n} $

\smallskip $ = \frac{v_D}{(w+1)^n} \int \limits_0^1 dx \cdot x^a \cdot \sum \limits_{k=0}^\infty \frac{x^k}{k!} \sum \limits_{l=0}^{k/2} \frac{(n+k-l-1)!k!}{(n-1)!} \frac{c^l}{l!} \frac{b^{k-2l}}{(k-2l)!} $

\smallskip $ = \frac{v_D}{(w+1)^n} \sum \limits_{k=0}^\infty \sum \limits_{l=0}^{k/2} \frac{(n+k-l-1)!}{(n-1)!} \frac{c^l}{l!} \frac{b^{k-2l}}{(k-2l)!} \int \limits_0^1 dx \, x^{a+k} $

\smallskip $ = \frac{v_D}{(w+1)^n} \sum \limits_{k=0}^\infty \sum \limits_{l=0}^{k/2} \frac{(n+k-l-1)!}{(n-1)!} \frac{c^l}{l!} \frac{b^{k-2l}}{(k-2l)!} \frac{1}{a+k+1} $

\bigskip
\noindent
with 

\bigskip $ v_D^{-1} = 2^{D+1} \pi^{D/2} \Gamma(D/2)$

\medskip $ a = m+\frac{D}{2}-1 $

\medskip $ b = \frac{1-\z}{1+w} $

\medskip $ c = -\frac{\zz}{1+w} $

\bigskip
\noindent
The series expansion and reordering is valid and convergent as long as
$ 1-bx-cx^2 \neq 0 $ for $ x \in [0;1] $. Which is the case, as the effective average propagator stays finite at all times.

\newpage
\subsection*{Critical exponents}

\noindent
Starting with a quartic potential $ u = \frac{1}{2}(\rho - \kappa_\Lambda)^2 $ 
the flow equations are discretized on a grid and numerically solved for different values of 
$ \kappa_\Lambda $. During the evolution towards $ k \to 0 $, the running minimum of the
potential (~$u^\prime(\kappa)=0$~) may either end up in the massless spontaneously broken phase 
(~$m=0$, $\kappa \to \infty $~) or in the massive symmetric phase (~$ m^2 \sim k^2 u^\prime(0)$, $\kappa = 0 $~). 

\bigskip
\noindent
The critical $ \kappa_c $ leads to a second order phase transition characterized by fix points 
of all couplings and vanishing masses. Fine tuning $ \kappa_\Lambda $ around $ \kappa_c $ then
yields the critical exponent $ \nu $ through the scaling law $ m \sim | T-T_c |^\nu $
and the established proportionality $ | \kappa_\Lambda - \kappa_c | \sim | T - T_c | $.
The critical exponent $ \eta $ may be identified with the fix point $ \eta^* $ of the anomalous 
dimension. 

\bigskip
\noindent
The critical exponents for the universality class of the three dimensional Ising model
obtained in this way are compared in Table 1 to values obtained in 
leading order (~$Z_1=0$, $Z_0=1$, also known as 'local potential approximation'~),
lowest order beyond leading order (~$Z_1=0$, $Z_0=const$, here called 'lowest order'~),
and full first order beyond leading order (~$Z_1=0$, here called 'first order'~).

\bigskip
\noindent
The exponents are in agreement with former calculations using the effective average
action. For brevity we quote only results which are based on simulations as similar
as possible. Still they differ in that they either use a field expansion ( an expansion
of $u$ and $z$ in powers of $\rho$ ) or a different infrared regulator or both. 
This might explain the numerical differences.

\bigskip
\noindent
For comparison we also give results from different approaches. 
An extensive review of different techniques and recent results can be found in \cite{Lit7}.
We adopt their overall estimates of exponents and errors. In addition we quote some of the 
newest and precisest results from high temperature expansions, perturbation series at fixed 
dimension, $\epsilon$-expansions and Monte Carlo simulations plus some additional references.

\subsection*{Conclusion and outlook}

\noindent
The critical exponent $\nu$ has significantly improved compared to 
first order. This is the first calculation using a linear cutoff that is consistent with
high precision methods. On the other hand, the anomalous dimension is sensitive to the 
neglected contributions as can be seen in comparison to \cite{Lit5a}.
Its error has even increased in comparison to first order. \\
Hence the inclusion of the neglected wave function renormalizations should be the next step.

\newpage
\onecolumngrid
\begin{center}
\begin{tabular}{|l|c|c|}
\hline
~&$\nu$&$\eta$\\
\hline
$~~~$effective average action to leading order ( this paper, $Z_1=0$, $Z_0=1$ )&0.6491&0\\
$~~~$effective average action to lowest order ( this paper, $Z_1=0$, $Z_0=const$ )&0.6271&0.1120\\
$~~~$effective average action to first order ( this paper, $Z_1=0$ )&0.6255&0.0503\\
$~~~$effective average action beyond first order ( this paper ) &0.6303&0.0793\\
\hline
$~~~$effective average action to leading order ( field expansion \cite{Lit6d} )&0.6496&0\\
$~~~$effective average action to lowest order ( field expansion \cite{Lit6e} )&0.6262&0.1147\\
$~~~$effective average action to first order ( exponential cutoff \cite{Lit7a} )&0.6307&0.0470\\
$~~~$effective average action to second order ( exp. cutoff, field exp. \cite{Lit5a} ) &0.632&0.033\\
\hline
$~~~$literature values from the review \cite{Lit7}&0.6301(4)&0.0364(5)\\
\hline
$~~~$25th-order high-temperature expansion \cite{HighT1}, see also \cite{HighT2}&$~~~$0.63012(16)$~~~$&$~~~$0.03639(15)$~~~$\\
$~~~$seven-loop perturbation series at fixed D=3 \cite{dExp}, see also \cite{eExp}$~~~$&0.6303(8)&0.0335(6)\\
$~~~$five-loop order $\epsilon$ expansion with boundary conditions \cite{eExp}&0.6305(25)&0.0365(50)\\
$~~~$Monte Carlo simulation \cite{Monte1}, see also \cite{Monte2,Monte3}&0.6297(5)&0.0362(8)\\
\hline
\end{tabular}

\bigskip
\noindent
TABLE 1: Critical exponents of the three dimensional Ising universality class
\end{center}
\twocolumngrid

\bigskip
\noindent
~


\begin{thebibliography}{10}

~

\bibitem{Lit1}
J. Berges, N. Tetradis and C. Wetterich,\\
Phys. Rept. {\bf 363} (2002) 223-386 $[$hep-ph/0005122$]$.

~

\bibitem{Lit2}
C. Wetterich,\\
Nucl. Phys. B {\bf 352} (1991) 529.

~

\bibitem{Lit3}
C. Wetterich,\\
Z. Phys. C {\bf 57} (1993) 451.

~

\bibitem{Lit4}
C. Wetterich,\\
Phys. Lett. B {\bf 301} (1993) 90.

~

\bibitem{Lit5}
N. Tetradis, C. Wetterich,\\
Nucl. Phys. B {\bf 422} (1994) 541-592 $[$hep-ph/9308214$]$.

~

\bibitem{Lit5a}
L. Canet, B. Delamotte, D. Mouhanna, J. Vidal,\\
$[$hep-th/0302227$]$.

~

\bibitem{Lit6}
D. F. Litim,\\
Phys. Rev. D {\bf 64} (2001) 105007 $[$hep-th/0103195$]$.

~

\bibitem{Lit6a}
D. F. Litim,\\
Phys. Lett. B {\bf 486} (2000) 92-99 $[$hep-th/0005245$]$.

~

\bibitem{Lit6b}
D. F. Litim,\\
JHEP {\bf 0111} (2001) 059 $[$hep-th/0111159$]$.

~

\bibitem{Lit6c}
D. F. Litim,\\
Int. J. Mod. Phys. A {\bf 16} (2001) 2081 $[$hep-th/0104221$]$.

~

\bibitem{Lit6d}
D. F. Litim,\\
Nucl. Phys. B {\bf 631} (2002) 128-158 $[$hep-th/0203006$]$.

~

~

~

~

~

~

~

\bibitem{Lit6e}
D. F. Litim,\\
private communication.

~

\bibitem{Lit7}
A. Pelissetto, E. Vicari,\\
Phys. Rept. {\bf 368} (2002) 549-727 $[$cond-mat/0012164$]$.

~

\bibitem{Lit7a}
S. Seide, C. Wetterich,\\
Nucl. Phys. B {\bf 562} (1999) 524 $[$cond-mat/9806372$]$.

~

\bibitem{HighT1}
M. Campostrini, A. Pelissetto, P. Rossi, E. Vicari,\\
Phys. Rev. E {\bf 65} (2002) 066127 $[$cond-mat/0201180$]$.

~

\bibitem{HighT2}
P. Butera, M. Comi,\\
Phys. Rev. B {\bf 65} (2002) 144431 $[$hep-lat/0112049$]$.

~

\bibitem{dExp}
F. Jasch, H. Kleinert,\\
J. Math. Phys. {\bf 42} (2001) 52 $[$cond-mat/9906246$]$.

~

\bibitem{eExp}
R. Guida, J. Zinn-Justin,\\
J. Phys. A {\bf 31} (1998) 8103 $[$cond-mat/9803240$]$.

~

\bibitem{Monte1}
M. Hasenbusch,\\
Int. J. Mod. Phys. C {\bf 12} (2001) 991.

~

\bibitem{Monte2}
H. W. J. Bl\"ote, L. N. Shchur, A. L. Talapov,\\
Int. J. Mod. Phys. C {\bf 10} (1999) 1137 $[$cond-mat/9912005$]$.

~

\bibitem{Monte3}
H. G. Ballesteros, L. A. Fern\'andez, V. Mart\'{\i}n-Mayor,\\A. Mu\~noz Sudupe, G. Parisi, J. J. Ruiz-Lorenzo,\\
J. Phys. A {\bf 32} (1999) 1 $[$cond-mat/9805125$]$.

\end{thebibliography}
\end{document}